\documentclass[twocolumn,pre,letterpaper,showpacs]{revtex4}
\usepackage{amsbsy}
\usepackage{amssymb}
\usepackage[dvips]{graphicx}

\usepackage{amsthm}
\usepackage{amsmath}
\usepackage{amsfonts}
\usepackage[compact]{titlesec}
\usepackage[utf8]{inputenc}
\usepackage{bbm}
\usepackage{enumerate}
\usepackage{setspace}
\usepackage{array}
\usepackage{listings}

\begin{document}
\title{Return probabilities and hitting times 
of random walks on sparse Erd\"os-R\'enyi graphs}

\author{O. C. Martin}
\affiliation{Univ Paris-Sud, LPTMS ; CNRS, 
UMR8626, Orsay, F-91405, France.}

\author{P. \v{S}ulc}
\affiliation{Univ Paris-Sud, LPTMS ; CNRS, 
UMR8626, Orsay, F-91405, France.}

\date{\today}

\begin{abstract}
We consider random walks on random graphs, focusing on
return probabilities and hitting times
for sparse Erd\"{o}s-R\'{e}nyi graphs.
Using the tree approach which is expected to be exact in the
large graph limit, we show how to solve for the \emph{distribution} of these
quantities and we find that these distributions exhibit a form of self-similarity.
\end{abstract}

\maketitle

\section{Introduction}
\label{sect:introduction}

Random walks are some of the simplest stochastic 
processes~\cite{Hughes96,BouchaudGeorges90} and yet they
arise in many scientific fields such as pure mathematics,
statistical physics or 
even biology~\cite{MontrollShuler79,Newman05,NohRieger04,BenichouCoppey05}. 
A fundamental quantity for computing properties
of random walks is the first passage time~\cite{Feller50,Redner01}.
Consider a random walk on a graph $G$, starting at 
node $s$; given another arbitrary node $t$ (the target), the hitting 
time $H(s,t)$ is just the mean of the 
first passage time to go from $s$ to $t$.
There is a well known relation between the value of $H(s,t)$ averaged
over all nodes $t$ of the graph and the spectrum of its 
adjacency matrix, as derived in~\cite{Lovasz93}.  

In this work we focus on random graphs~\cite{Bollobas01,Lovasz93}.
For dense Erd\"{o}s-R\'{e}nyi graphs~\cite{ErdosRenyi59},
the spectrum of the diffusion operator converges to that of 
a Gaussian random matrix
and one can show~\cite{ChungLu2003,SoodRedner05} that if $N$ 
is the number of nodes of $G$,
the hitting time is $N + o(N)$.
As far as we know, there is no analogous result for sparse graphs:
only a mean-field approximation has been 
derived~\cite{BaronchelliLoreto06} which neglects certain
fluctuations. This situation is surprising because
the problem has been open for many years, but
the lack of progress underlies the 
difficulty of deriving
analytically the spectrum of the adjacency
matrix on sparse random graphs~\cite{BiroliMonasson99,semerjian}.
Nevertheless, we here
bypass this difficulty by exploiting 
the local structure of sparse random graphs that
is tree-like with probability $1$ at large $N$. 
If, as in a number of other problems~\cite{Aldous01,MezardParisi02},
only the graph's local structure matters at large $N$,
then the problem maps in the
$N \to \infty$ limit to diffusion 
processes on random trees. This tree approach, which will be validated
in Sect.~\ref{sect:num}, then provides an 
analytical calculation for the hitting times and for a 
closely related quantity, the
probability that the walker returns to its starting node in a finite time.

In what follows, we first specify the stochastic dynamics
of the random walk and the kinds of random graphs we use. After
we compute the hitting times and probabilities of return
on random $d$-regular graphs~\cite{Wormald99}.
That calculation is then generalized to sparse Erd\"os-R\'enyi graphs,
displaying quite subtle distributions.


\bigskip

\section{The model}
\label{sect:model}

We consider a random walker on a graph $G$. At each time step
$n$, the walker hops to one of the neighboring
nodes, all such nodes being equi-probable. 
It is convenient to introduce
the adjacency matrix $A$ of $G$: $A_{ij}=1$ if nodes $i$ and $j$ 
are connected by an edge and $A_{ij}=0$ otherwise. Defining at each time step $n$
the probability $\mathbf{v}_i^{(n)}$ of having the walker be at
node $i$, the vector of probabilities obeys the 
master equation
\begin{equation}
\label{stomatice}
  \mathbf{v}_i^{(n+1)}  = 
\sum_{<ji>} \frac{1}{d_j} \mathbf{v}_j^{(n)} = \left(  A D^{-1} \mathbf{v}^{(n)} \right)_i
\end{equation}
where the sum is taken over all nodes $j$ that are adjacent to the node
$i$. The matrix $D$ is diagonal; its $i$-th diagonal element $D_{ii}$ is equal 
to the degree $d_i$ of the
$i$-th node.

To investigate the hitting time of the walker to go from 
node $s$ to $t$, it is enough to initialize the vector $\mathbf{v}^{(0)}$ to
be zero on all nodes except at $s$ where it is 1, and to impose
absorbing conditions at the target node $t$, \emph{i.e.}, 
$\mathbf{v}_t^{(n)}=0$ at
all $n$. Then the probability of having a first passage time 
equal to $n$ 
is given by the flux into node $t$ at that 
time step~\cite{Feller50}. A modified
treatment of the walker allows one to also obtain 
the probability of return to the starting node.

Our mathematical solution concerns
Erd\"{o}s-R\'{e}nyi graphs in the ensemble $G(N,p)$, where $N$ is the 
total number of nodes and each pair of nodes has probability $p$ to be 
connected by an edge. For sparse 
graphs, $p=c/N$ where $c=\langle d \rangle$ is the 
mean degree of nodes.
We shall also consider fixed degree random graphs, also called
random $d$-regular graphs, where each node has the same degree $d$
and connections are otherwise random~\cite{Wormald99}.

\bigskip

\section{Hitting times on random $d$-regular graphs}
\label{sect:hitting_cayley}
Let us first compute the hitting time on random regular graphs,
exploiting their local
tree-like nature. Clearly, 
loops can arise in random $d$-regular graphs~\cite{Wormald99}
but their typical length is $O(\ln(N))$. Thus it is expected that
most properties can be obtained by studying what happens locally,
as long as boundary conditions at ``infinity'' are properly
handled. Such an approach has been used in many contexts
with a high level of success~\cite{Aldous01,MezardParisi02}.

For a given random regular graph, of fixed degree $d$,
we consider a node $t$ and ask what is the mean of $H(s,t)$
when averaged over all possible departing nodes $s$. We need to solve a
diffusion problem where at time $n=0$ a walker is equi-distributed
amongst the $N-1$ nodes $s$ ($s \ne t$) and if the walker
hits node $t$ it gets absorbed.
If one denotes
by $F_t^{(n)}$ the probability flux into node $t$ at step $n$, then
the hitting time averaged over all $s$ is given by the first
moment of $n$ distributed according to $F_t^{(n)}$.

In the neighborhood of $t$, the graph is a
Cayley tree with probability one at large number of nodes $N$ and thus
does not depend on the node which we choose as absorbing in the large $N$ limit.
Given the diffusion-absorption process,
the vector of probabilities quickly converges to
the dominant eigenvector of the master equation (that
with the largest eigenvalue, decaying the slowest). In the limit
of large $N$, the decay rate goes to zero and all the transient
behavior (associated with the other eigenvectors) becomes irrelevant.
When $N \to \infty$, it is then enough to determine
the dominant eigenvector, imposing zero boundary
condition at the root node $t$ 
and $1/(N-1)$ boundary conditions for the 
far away nodes.

As $N\to \infty$, the recurrence equation that is satisfied by the
eigenvector's elements leads to
$d A_{k+1} = A_{k+2} + (d-1) A_k$
where $A_k$ is the sum of the probabilities on the nodes that are at
distance $k$ from the root node.
Solving this, subject to the normalization
and boundary conditions, leads to the value
of $A_1$ and thus the flux flowing into the 
absorbing node using the eigenvector: $F_t = A_1/d$.

Note that since at large $N$ only the leading eigenvector matters,
the first passage time is
exponentially distributed with a mean given by the inverse of this flux.
This then gives for random $d$-regular graphs 
a hitting time behaving at large $N$ as
\begin{equation}
\label{rrgraph}
\frac{H}{N} = \frac{d-1}{d-2}  + o(1) \ .
\end{equation}
Finally, it is worth noting that for random $d$-regular graphs, with probability 1 in 
the large $N$ limit, the ratio
$H(s,t)/N$ does not depend on the starting node $s$. Also,
because of the regularity of the graph, this quantity
does not depend on $t$ either.

\bigskip

\section{Probability of return on random $d$-regular graphs}
\label{sect:return_cayley}

On any finite graph, a walker leaving node $t$ 
will return with probability
one. Nevertheless, if one considers the distribution
of return times for increasing values of $N$, one will find that 
there is a $N \to \infty$ limiting
point-wise distribution but which does not integrate to 1. Indeed, 
in that limit, the return times will be finite with 
probability $\hat{r}$ and will diverge linearly in $N$ with 
probability $1-\hat{r}$. If $\hat{r} \neq 1$, the walk is said to be transient.
On the infinite Cayley tree, $\hat{r}$ can be computed simply by using the
homogeneity of the graph as follows.

Take $t$ to be the root of an infinite Cayley tree. 
The walker must make a first step; let it be to one of its neighbors $j$.
Define $r$ as the probability for the walk to return to $t$ \emph{given}
that it has stepped to $j$. Using the equivalence of all
nodes, one can write a series for $r$:
\begin{equation}
\label{eq:r_series}
r = \frac{1}{d} + \frac{(d-1) r}{d} \frac{1}{d} +  \frac{((d-1) r)^2}{d^2} \frac{1}{d} + \ldots
\end{equation}
where $d \ge 2$ is the degree of the Cayley tree. In this series,
the term of $O(r^p)$ corresponds to the probability that the walk 
returns $p$ times to node $j$ before going back to the
root $t$. Summing this geometric series gives two possible values:
$r = 1$ and $r = 1 / (d-1)$. Furthermore, it is easy to see that $\hat{r} = r$. If $d=2$, we have a one dimensional walker and $\hat{r}=1$. For $d \geq 3$, the walk is transient and $\hat{r} = 1/(d-1)$.

\bigskip 

\section{Probability of return on Erd\"os-R\'enyi graphs}
\label{sect:return_ER}

Here we extend the previous calculation of return
probabilities to the case of Erd\"os-R\'enyi graphs.
Just as for the random $d$-regular graphs, we exploit the fact that
with probability $1$ in the large $N$ limit
the neighborhood of a node belonging to a sparse Erd\"os-R\'enyi graph
is locally tree-like. We denote by $c=\langle d \rangle$ the mean degree
of these graphs; the probability to have a node of
degree $d$ is
$P(d)= e^{-c} c^d / d!$, \emph{i.e.}, is given by the Poisson distribution.

To find the probability to return in a finite number of steps 
(formally at infinite $N$) 
for a walker starting on the root node $t$, we reconsider the 
series of Eq.~\eqref{eq:r_series}. Suppose that at the first step
the walker moves to the neighbor $j$ of the root node, and that
$d_j$ is the connectivity of that node.
If the walker is to return to $t$, it can do so immediately, or 
it can perform $p$ loops from $j$ (avoiding $t$), 
stepping back to $t$ only
after its $(p+1)$th visit to node $j$.
By a loop from $j$, we mean a step to one of the $d_j-1$ 
neighbors of $j$ other than $t$, then a finite number of steps that do not
visit $j$, and then finally a return to $j$. The point is that 
in our system the walker cannot come back to $t$ other than through the
edge connecting $j$ to $t$: any other route requires going to
``infinity'' and thus an infinite number of steps. (Since we
are dealing with the return probability on an infinite graph,
the walks returning to $t$ must have a finite number of steps.)

For the edges connecting node $j$ to a node
other than $t$, let the return probabilities be
$r_j(1)$, $r_j(2)$, \ldots $r_j(d_j-1)$. 
Given these $r_j$s, 
the probability $r$ to return to the root node
if the walk's first step is to node $j$ is
\begin{equation}
\label{eq:r_functional}
r = \frac{1}{d_j - \sum_{m=1}^{d_j-1} r_j(m)} \, .
\end{equation}
However, the $r_j(m)$ are \emph{i.i.d.} random
variables belonging to a distribution $\rho(r)$.
In the Erd\"os-R\'enyi ensemble, $t$ connects to a random 
node ($j$ here) which itself connects to other random nodes.
The distribution of $r$ is thus the same as that of the 
$r_j$s, and Eq.~\eqref{eq:r_functional} determines implicitly
a self-consistent functional equation for $\rho(r)$. This can 
be written formally as:
\begin{eqnarray}
\label{eq:r_functional_long}
\rho(r) =  P(0) \delta(1 -r) &+&  \sum_{z=1}^{\infty} P(z) \int dr_1  \ldots  \int dr_z  \\
\rho(r_1) \ldots \rho(r_z) 
 &\times & \delta \left( \frac{1}{1 + \sum_{i=1}^{z} (1-r_i)} - r \right) \nonumber
\end{eqnarray}
where $P(z)$ is the Poisson distribution (of $z=d_j-1$), and
$\delta(x)$ is the Dirac delta function. Note that since we are 
dealing with an Erd\"os-R\'enyi graph, the probability that the node $j$ 
(which by construction is connected to the absorbing node $t$)
has degree $z+1$ is given by $P(z)$. This is due to the 
fact that for Erd\"os-R\'enyi graphs, the edges are independent.

\begin{figure}[t]
 \begin{center}
 \includegraphics[width=8cm, height=5cm, angle=0]{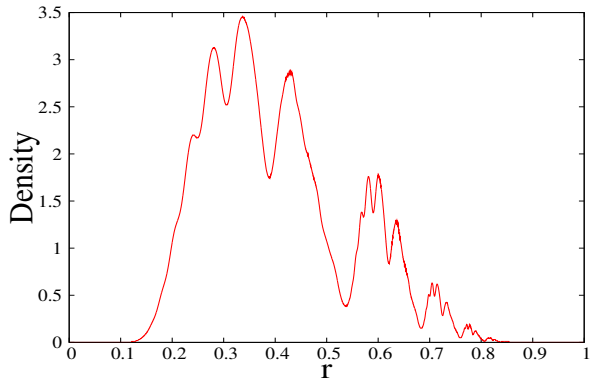}
\end{center}
 \caption{(Color online) The probability density of the return probability $r$ 
after stepping from a given node to one of its neighbors on an infinite
Erd\"os-R\'enyi graph with mean degree 3. For ease of presentation,
the delta function contribution at $r=1$ has been removed and
the rest has been rescaled to have a total probability of 1.
Note the qualitative self-similarity.
}
 \label{fig:r_distribution}
\end{figure}

We have solved for $\rho$ by numerical iteration, demanding a
stable distribution. Because $\rho$ has both a continuous
part for $0 \leq r < 1$ and a delta function part at $r=1$,
it was necessary to treat these two parts separately, and the
the convergence in the number of iterations
is quite fast. To illustrate our results,
we display in Fig.~\ref{fig:r_distribution} the probability density $\rho(r)$ when
the mean degree is $3$.
(Numerically, we must introduce a
coordination cut-off and binning to compute $\rho(r)$; we find that taking a 
cut-off value of a few times the graph's mean coordination leads to negligible errors, while
beyond 2500 bins no visible dependence on bin size can be seen. For all the
figures presented here, we used $10000$ bins.)
It also exhibits a form of self-similarity:
the motif for $0 < r < 0.5$
is repeated at larger values of $r$ but each time with a smaller
amplitude and some distortion. 
Also, note that the distribution is relatively 
smooth; its continuity can be justified as follows. 
Consider the ensemble of graphs for which the return probability $r$ is in the interval 
$[r, r+dr]$. If we increase slightly the degree of a 
node far away from the absorbing node
for all of these graphs, the return probability $r$ will decrease slightly. If this modified node is 
sufficiently far, the change in $r$ can be made arbitrarily small. Because of this, the distribution of $r$ 
can have no discontinuities.

As a last point, the intensity $\Delta$ of the Dirac part
of $\rho$ gives the probability for the first step of the walk
to connect to a finite part of the graph. It is thus 
simply given~\cite{Bollobas01} by
the solution to the equation
$\Delta =  \sum_{k=0}^{\infty} P(k) \Delta^k$, obtained by forcing
the node $j$ to have all its neighbors in a finite part of the graph also. 
In such a situation, one has $r=1$.

\bigskip

\section{Hitting times on Erd\"os-R\'enyi graphs}
\label{sect:hitting_ER}

To compute the hitting time $H(s,t)$,
we take $s$ and $t$ to be on the same connected component
whose size we denote by $N_{\infty}$.
For Erd\"os-R\'enyi graphs, we work beyond the
percolation threshold, $c > 1$, on the ``infinite''
component, so $N_{\infty} \simeq (1-\Delta) N$. 
With probability
$1$, the hitting time $H(s,t)$ scales with $N$, 
has negligible fluctuations
with $s$, and depends only the neighborhood properties
of $t$. We thus focus on $H_t$, the mean of $H(s,t)$ when
averaging over all nodes $s$ distinct from $t$. This problem
has been solved for \emph{dense} Erd\"os-R\'enyi graphs
and leads to $H_t = N + o(N)$~\cite{SoodRedner05}.
For the sparse case, no exact treatment has been proposed, 
but a mean-field like approximation gives
rather good results~\cite{BaronchelliLoreto06}. We now provide
an exact mathematical approach.

As explained previously, we can follow the probability of finding
the walker on any node. 
The initial condition is that every node
except $t$ is occupied
with the same probability $1/(N_{\infty}-1)$.
The absorption at node $t$ imposes
$\mathbf{v}^{(n)}_t=0$ at all times. The master equation for this process is 
therefore
\begin{equation}
\label{absmaster}
 \mathbf{v}^{(n+1)}  = \left( T A D^{-1} \mathbf{v}^{(n)} \right)
\end{equation}
where $T_{ij} = \delta_{ij} (1 - \delta_{ti})$.
\begin{figure}[t]
 \begin{center}
 \includegraphics[width=8cm, height=5cm, angle=0]{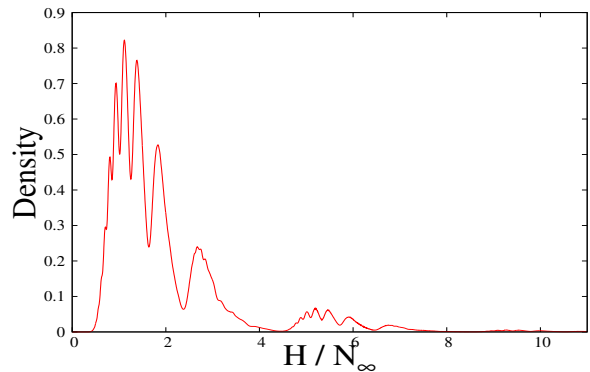}
\end{center}
 \caption{(Color online) The probability density of $H/N_{\infty}$ 
on Erd\"os-R\'enyi graphs with mean degree 4, in the large
graph size limit. $H$ is the hitting time of walks 
residing on the graph's
infinite (percolating) component and absorbed at a random
node $t$; $N_{\infty}$ is the size of that connected component.}
 \label{fig:H_distribution}
\end{figure}
Denote by $\mathbf{S}$ the leading eigenvector of the 
diffusion operator $A D^{-1}$ having \emph{no absorption}, with eigenvalue $1$. For
a normalisation of the probabilities to 1, one has
$\mathbf{S}_i = d_i / ( N_{\infty} \langle d \rangle_{\infty})$
where $d_i$ is the degree of node $i$ on the infinite component. 
Furthermore, $\langle d \rangle_{\infty}$ is the mean degree on the
connected component considered, which in our case is
not $c$ because we have the constraint of belonging to the
infinite component, instead it is
\begin{equation}
\langle d \rangle_{\infty} = \frac{\sum_{k=1}^{\infty} 
k (1 - \Delta^k) P(k)}
{\sum_{k=1}^{\infty}  (1 - \Delta^k) P(k) }  .
\end{equation}
It is easy to check that under evolution without absorption $\mathbf{S}$ is 
unchanged: since the walk is on a connected component,
this is the only normalized steady state distribution. Now 
introduce the vector
$\mathbf{b}^{(n)}$ that represents the difference 
between the vector $\mathbf{S}$
and the vector $\mathbf{v}^{(n)}$ :
\begin{equation}
 \label{becko}
\frac{1}{N_{\infty}} \mathbf{b}^{(n)}_i = \frac{1}{N_{\infty}}
       \frac{d_i}{\langle d \rangle}_{\infty}  - \mathbf{v}^{(n)}_i.
\end{equation}
The absorption condition at $t$ then imposes
$\mathbf{b}^{(n)}_t =  d_t / \langle d \rangle_{\infty}$ for all $n$.
Far away from the root node, the distribution quickly relaxes
to the leading eigenvector of the diffusion equation. In
the $N_{\infty} \to \infty$ limit, almost all nodes are oblivious to
the absorption, so we can compute the hitting time 
by assuming that $\mathbf{v}^{(n)}_m$ is equal to $\mathbf{S}_m$ for all nodes $m$ at ``infinity'', which gives us 
the boundary condition $\mathbf{b}^{(n)}_m = 0$ at all times.

Now we can interpret the evolution equation for $\mathbf{b}^{(n)}$ as describing
a process of multiple random walkers diffusing on the graph, with
in addition a fixed source at the root node. Specifically,
at each time step $n$, 
$\mathbf{b}_t^{(n)}$ new walkers are created at the root and step away while any walkers incoming to the root are removed from the system.
With increasing number of iterations, the vector 
$\mathbf{b}^{(n)}$ converges to a steady-state $\tilde{\mathbf{b}}$ (as $\mathbf{v}^{(n)}$ converges to $\tilde{\mathbf{v}}$, a leading eigenvector of $TAD^{-1}$) in which 
for each edge $\langle tj\rangle$ connected to the root node, there is
an outgoing flux of $1/\langle d \rangle_{\infty}$ and 
a corresponding incoming flux of $r_j/\langle d \rangle_{\infty}$ where $r_j$ is the
probability of return to $t$ of a walker given that it has stepped to $j$. The flux into $\tilde{\mathbf{b}}_t$ is then equal to the flux of ``returning'' random walkers:
\begin{equation}
\label{bflux}
 \sum_{<jt>} \frac{1}{d_j} \tilde{\mathbf{b}}_j = \frac{1}{ \langle d \rangle_{\infty}} \sum_{<jt>} r_j .
\end{equation}
Coming back to the formalism based on $\tilde{\mathbf{v}}$, \emph{i.e.}, the leading eigenvector of $TAD^{-1}$, the \emph{net total} 
flux $F_t$ into the absorbing node $t$ is given by
\begin{equation}
\label{zeroflux}
  F_t = \sum_{<jt>} \frac{1}{d_j} \tilde{\mathbf{v}}_j .
\end{equation}
Using Eqs.~\eqref{becko} and \eqref{bflux} one obtains the final expression
\begin{equation}
\label{retprob}
F_t = \frac{1}{N_{\infty} \langle d \rangle}_{\infty} \sum_{<jt>} \left(1 - r_j \right) .
\end{equation}
In the previous section we derived the distribution of $r_j$; from
that we easily obtain the distribution for $H_t = 1/F_t$ as follows.
First, for each value of $d_t$ (the degree of the root node), we compute the
distribution of $F_t$. The delta function part of this distribution (at $F_t
= 0$) is removed and the remaining distribution if rescaled to have norm
1. This corresponds to enforcing the constraint that the absorbing node is
on the infinite component of the Erd\"os-R\'enyi graph (the part of the
distribution of $F_t$ which gives zero flux corresponds to being on a finite
component). Second, the distribution of $H_t = 1/ F_t$ is extracted: call 
it $\mu_{d_t}(H_t)$. Finally, given all the distributions $\mu_{d_t}$ ($1
\leq d_t < \infty$), the distribution of hitting times $H$ 
at \emph{random nodes} is obtained by averaging the $\mu_{d_t}$ with their respective weights:
\begin{equation}
  \label{hacko}
 \mu( H ) = \sum_{d_t=1}^{\infty} \frac{  \mu_{d_t} \left( H_t \right) P(d_t) (1-\Delta^{d_t}) }{\sum_{j=1}^{\infty}  P(j) (1-\Delta^j) } 
\end{equation}
An example of such a distribution is shown in Fig.~\ref{fig:H_distribution} when $\langle d \rangle = 4$.
Furthermore, the distribution of $H$ also gives the distribution
of first passage times since at large $N$, for each value of $H$,
the first passage time $n$ is distributed as $\exp{(- n / H)}$.
\begin{figure}[t]
 \begin{center}
 \includegraphics[width=8cm, height=5cm, angle=0]{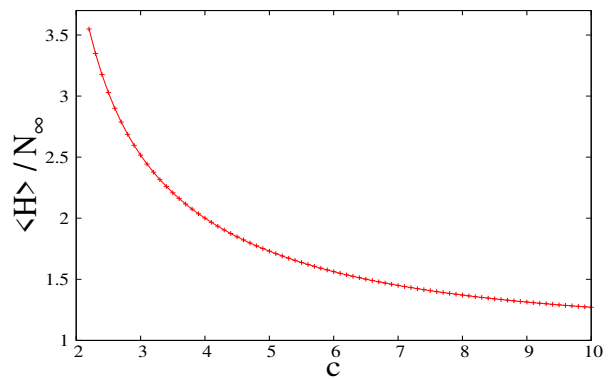}
\end{center}
 \caption{(Color online) Mean hitting times divided by $N_{\infty}$ 
for Erd\"{o}s-R\'{e}nyi graphs in the limit of large graphs,
as a function of mean node degree $c=\langle d \rangle$.
$N_{\infty}$ is the size of the ``infinite'' component,
$N_{\infty} \approx (1-\Delta)N$ for graphs of $N$ nodes.}
\label{fig:d_dependence}
 \end{figure}
Finally, to obtain the \emph{mean} hitting time $\langle H \rangle$, it is 
enough to compute the mean of the distribution of $H$. 
We have done so and show in Fig.~\ref{fig:d_dependence}
the resulting values, normalized by $N_{\infty}$,
as a function of the mean degree of the graphs.
At large $\langle d \rangle $, the ratio converges to $1$ with $O(1/ \langle d \rangle)$ corrections:
one recovers the  
dense graph result. Also, the behavior is very
smooth and we find that it differs from the value when the degree
does not fluctuate (the case of random $d$-regular graphs) also by
$O(1/\langle d \rangle)$. 

\bigskip
\section{Validation of the tree approach}
\label{sect:num}
%
\begin{figure}[t]
 \centering
 \includegraphics[width=8cm, height=5cm, angle=0]{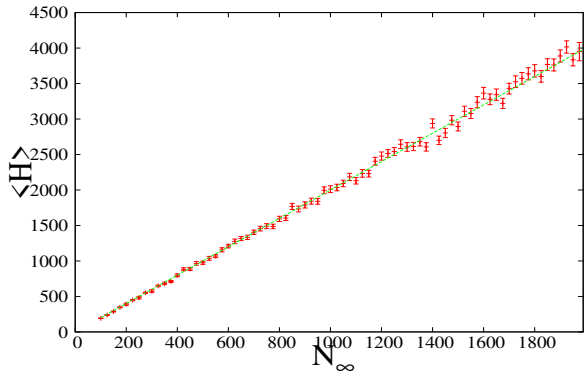}
\caption{(Color online) Plot comparing numerical simulation with analytical results. The x axis shows the size of the largest connected component of the graph, the y axis shows the mean hitting time for such a component.}
\label{fig:ebar}
\end{figure}

One of the key assumptions in the derivation of our formulas is
that, since the graphs under consideration are locally tree-like, 
quantities such as the return probability can 
be computed by replacing the graphs by trees with the same 
statistics for the node degrees.
There are certain systems where such an approach can be 
demonstrated to be exact in the large graph
limit~\cite{Aldous01}, but unfortunately in most cases one
has no such a proof. To see whether the
tree approach might be exact (for large graphs) for the 
mean hitting times, we have computed by simulation 
the actual values for random graphs without resorting to any approximation.
These values can then be compared to the theoretical predictions,
in particular in the large graph size limit.

Fig.~\ref{fig:ebar} shows the mean hitting times on the largest 
connected component of an Erd\"{o}s-R\'{e}nyi graph with mean 
degree $\langle d \rangle = 4$.
The estimation from Eq.~\eqref{hacko} (based on the
tree approach) is compared with values 
obtained from a numerical simulation in which we followed the 
probability vector $\mathbf{v}^{(n)}$ as in Eq.~\eqref{absmaster}. For each
randomly generated graph of size $N$, we numerically calculated the mean
hitting time for a randomly chosen absorbing node $t$ on 
its largest connected component (whose size is $N_{\infty}$). The mean
hitting times were then averaged over multiple graphs. The error bars are shown as well.
We found that the values $\langle H \rangle / N_{\infty}$ 
determined from the simulations tend towards their
large $N$ limit 
rather fast and that this limit is compatible with our analytical result,
the relative difference being
compatible with a $O(1/N)$ convergence. The same conclusion also holds 
in the context of random $d$-regular graphs (cf. Eq.~\eqref{rrgraph}).
In sum, the agreement of the 
theoretically predicted values with the results from numerical simulations
gives some credence
to the claim that the tree approach is exact in the large $N$ limit.

\bigskip

\section{Discussion and conclusion}
\label{sect:conclusions}

We considered random walks on random graphs, focusing on 
two quantities: 
the distribution of hitting times
and the probability that a walker will return
to its starting point in a finite time. (The hitting time is the mean
of first passage times.)
By using the local tree approach~\cite{Aldous01,MezardParisi02},
we were able to calculate analytically the large $N$ behavior
of these quantities on two families of random graphs.
We found non-trivial
distributions having self-similar features associated with the
discrete nature of possible neighborhoods of a node. 
Finally, we compared the calculated results with numerical simulations and
found excellent agreement, justifying the tree approach which assumes
that the loops in these graphs can be treated by appropriate boundary
conditions on infinite trees.

\bigskip

{\bf Acknowledgments---}
We thank Satya Majumdar for enlightening discussions.
This work was supported by the Sixth European Research 
Framework (proposal number 034952, GENNETEC project).
The LPTMS is an Unit\'e de Recherche de
l'Universit\'e Paris-Sud associ\'ees au CNRS.

\bibliographystyle{apsrev}
\bibliography{graphs,references}

\end{document}